\documentclass[submission,copyright,creativecommons]{eptcs}
\providecommand{\event}{INFINITY 2010} 
\usepackage{color,graphics,times,amsmath,amssymb}
\def    \s      {\sigma}
\def    \S      {\Sigma}
\def    \o      {\omega}

\def    \to     {\rightarrow}

\def    \l      {\lambda}
\def    \d      {\delta}

\def    \ee     {\varepsilon}
\def    \D      {\Delta}

\def    \t      {\tau}
\def    \th     {\theta}

\def    \f      {\varphi}
\def    \a      {\alpha}

\def    \A      {{\cal A}}

\def    \T      {{\cal T}}

\def    \R      {{\cal R}}

\def    \ni     {\noindent}

\def    \se     {\subseteq}

\def    \lcc#1  {\langle\!\langle #1 \rangle\!\rangle}
\def    \rcc#1  {\langle #1 \rangle}
\def    \cc#1   {[#1]}

\def	\qed	{\hfill\hbox{\hskip 4pt
                \vrule width 3pt height 1pt depth 1pt
                \hbox{\vrule width 2pt height 3.5pt depth 1pt}}}
\newcommand\trans[1]{\stackrel{{\scriptsize#1}}{\longrightarrow}}

\def	\ds     {\displaystyle }
\def	\comment#1 {}

\def    \comm#1  {\begin{center} {\tt $<<<$ #1 $>>>$ } \end{center}}

\def    \dt      {\D}    

\def\GF{\hbox{\raise 1.6pt \hbox{$\varphi$}}}

\newcommand \ov[1]{\overline{#1}}
\newcommand\vectorl[1]{\mathbf{#1}}
\newcommand\ft[1]{\footnote{#1}}

\newcommand\vi{\vectorl{1}}

\newcommand\sem[1]{[\![#1]\!]}

\newcommand{\N}{\mathbb{N}}
\renewcommand{\R}{\mathbb{R}}

\renewcommand{\T}{\mathbb{T}}
\newcommand{\eps}{\epsilon}

\newtheorem{observation}{Observation}
\newtheorem{definition}{Definition}
\newtheorem{theorem}{Theorem}

\renewcommand{\ov}[1]{{\overline{#1}}}
\newcommand{\wq}{\ov{q}}

\newcommand{\tpr}{\pi}
\newcommand{\st}{s}
\newcommand{\supp}{S}
\newcommand{\ltq}{L^{\to\circ}}
\newcommand{\lfq}{L^{\circ \to}}
\renewcommand{\dt}{\mathcal{T}}

\begin{document}


\title{On Zone-Based Analysis of Duration Probabilistic Automata}  
%
%

\author{Oded Maler
\institute{{\sc cnrs-verimag}\\ University of Grenoble \\ France}
\email{Oded.Maler@imag.fr}
\and
Kim G.~Larsen
\institute{CISS and CS \\Aalborg University \\ Denmark}
\email{kgl@cs.aau.dk}
\and
Bruce H.~Krogh
\institute{Department of EC \\ Carnegie Mellon University \\ USA}
\email{krogh@ece.cmu.edu}
}
\def\titlerunning{Duration Probabilistic Automata}
\def\authorrunning{O.~Maler, K.G.~Larsen \& B.H.~Krogh}
\maketitle
\begin{abstract}
We propose an extension of the zone-based algorithmics for analyzing timed automata to handle systems where timing uncertainty is considered as \emph{probabilistic} rather than \emph{set-theoretic}. We study \emph{duration probabilistic automata} (DPA), expressing \emph{multiple} parallel processes admitting \emph{memoryfull} continuously-distributed  durations. For this model we develop an extension of the zone-based forward reachability algorithm whose successor operator is a \emph{density transformer}, thus providing a solution to verification and performance evaluation problems concerning acyclic DPA (or the bounded-horizon behavior of cyclic DPA). \comment{We also speculate about the convergence of this computation for unbounded horizon.}
\comment{From the point of view of stochastic processes, we propose a new computational technique for class of generalized semi-Markov processes.}
\end{abstract}

\section{Introduction}

Timed automata \cite{ad94} handle temporal uncertainty in a set-theoretic manner consistent with the \emph{worst-case} spirit of safety-critical verification. Performance evaluation of systems of a less dramatic nature is typically based on a stochastic interpretation of temporal uncertainty. A well-studied class of such systems are \emph{continuous-time Markov chains} (CTMC) where durations are distributed exponentially and model-checking against temporal properties is well understood \cite{baier-ctmc}. More general distributions fall under the category of \emph{generalized semi-Markov processes} (GSMP) \cite{glynn,german,deds-book} and other similar models such as \emph{stochastic timed automata} \cite{katoen1,modest} or \emph{stochastic Petri nets} \cite{spn-book1,spn-book2}. Good overviews of these issues can be found in \cite{BHK-book,bouyer-hab2009}.  Some approaches for verifying such systems against qualitative \cite{acd91} and quantitative \cite{kwiatkowska-concur2000}  properties have been proposed  based on partitioning the state space into equivalence classes in the spirit of the \emph{region graph} \cite{ad94} and performing the analysis on the finite quotient which can be viewed as a discrete-time Markov chain. Although the region graph underlies the fundamental decidability results for timed automata, it is not used in any existing verification tool, due to its prohibitive size. Verification tools \cite{kronos,uppaal} use  reachability computation on \emph{zones} \cite{hnsy}, a class of polyhedra that represent reachable sets of states and clock valuations.\ft{Theoretically the number of zones can be even higher than the number of regions but in practice it is much lower.}

We extend the zone-based reachability computation  to handle timed automata with probabilistic durations. We use a variant of stochastic timed automata that we call \emph{duration probabilistic automata} inspired by the class of timed automata encountered while modeling scheduling problems \cite{schedule-tcs}. Such automata can model tasks admitting precedences and resource constraints, with the duration of each task being probabilistically distributed. We focus on \emph{uniform} distributions but the proposed approach will work with any polynomial distributions with bounded support. To analyze such systems we  decorate zones with clock \emph{densities}, and define \emph{successor} operators that act as \emph{density transformers} that allow us to compute the clock distribution upon taking a particular transition from state $q$ based on the clock distribution at the entrance into $q$. As a result we can assign probabilities to interesting subsets of the \emph{timed language} generated by the automaton.

The rest of the paper is organized as follows. Section~2 is a self-contained introduction to the modeling of timing uncertainty in concurrent systems and its algorithmic analysis. In Section~3 we define duration-probabilistic automata. Section~4 is devoted to a summary of the reachability graph construction  used to compute the semantics of timed automata. In Section~5 we present our contribution, the extension of this technique for DPA using density transformers while Section~6 mentions related and future work.

\section{Timing Uncertainty: Modeling and Analysis}

Discrete concurrent processes can be analyzed at different levels of abstraction with respect to time. To illustrate this point consider two concurrent systems, one that performs two tasks sequentially and one that preforms a third task in parallel and let events $a$, $b$, and $c$ denote the respective \emph{terminations} of these tasks. At the most abstract level one assumes nothing about the relative \emph{durations} of the processes and hence all the sequences in the shuffle  $a b || c=\{abc,acb,cab\}$ are considered feasible. The first refinement of the model is provided by models such as \emph{timed automata} or \emph{timed Petri nets}, where the durations of $a$, $b$ and $c$ are specified to be bounded in the intervals $[l_a,u_a]$,  $[l_b,u_b]$  and $[l_c,u_c]$, respectively. In this model,  knowing, for example, that $l_c>u_a$ we  conclude that $c$ cannot occur before $a$ and hence $cab$ is impossible. Likewise,  $abc$ is impossible when $u_c<l_a+l_b$.

\begin{figure}
\begin{center}
\scalebox{0.5}{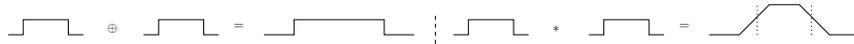 }
\end{center}
\caption{A Minkowski sum of intervals versus a convolution of two probability density functions. The area outside the dotted vertical lines corresponds to low-probability behaviors that can be ignored in certain circumstances. \label{fig:sum} }
\end{figure}

While this refinement of the untimed model adds a lot of  information, this \emph{set-theoretic nondeterminism} which states only \emph{what} is possible but does not quantify the likelihood of different possibilities,  is still too \emph{qualitative} for certain purposes as the following example demonstrates. Consider a sequence of $k$ processing steps, each of which with duration in $[l,u]$. From a purely ``measureless'' set-theoretic viewpoint, the termination time of the whole sequence of steps  can be anywhere in $[kl,k u]$.  Intuition tells us, however, that a duration of $k u$, whose realization requires that each of the steps  takes the maximal time to terminate, is less likely than, say, an ``average''  duration of $k(l+u)/2$.\ft{ Another example of a more discrete nature is the modeling of computer memory access where  worst case duration (cache miss) is orders of magnitude larger than the normal case (cache hit) and if we want to be conservative and assume that both cases are possible in each and every instance, our performance estimation will be overly pessimistic and practically useless. Timed automata with probabilities on transitions have been studied in \cite{jensen,kwiatkowska-discrete}.} On the other hand if we interpret the interval $[l,u]$ as, say, a uniform distribution with density $1/(u-l)$, the total duration of the $k$-step sequence is still restricted to the interval $[k l,k u]$, but with probability which is larger in the middle of the interval and smaller toward the boundaries. In a nutshell, this is the difference between a Minkowski sum of intervals $[l,u]\oplus [l,u]$ and the \emph{convolution}  $\psi_1*\psi_2$ of two functions defined over those intervals, see Fig.~\ref{fig:sum}. Assigning probabilities to the runs of the automaton we can, for example, distinguish between different degrees of property violations or compute the expected value  over all runs of some performance measure.

\comment{Augmenting timing information with this additional quantitative dimension allows one to go beyond the worst-case pessimistic reasoning underlying timed automata and verification in general.}

The use of automata with clocks has some advantages over the standard language of stochastic processes, in particular,  the ability to express more sophisticated synchronization mechanisms between processes, such as schedulers that resolve resource conflicts. These are expressed naturally in the language of states and transitions while translating them into conditionals based on inequalities over values of random variables may be cumbersome.
Computationally, a state-based  approach provides for iterative forward or backward computations, for both analysis and scheduler synthesis, more flexible than methods  based on a holistic analytical solution.\ft{We use \emph{automata} here as a generic term for discrete transitions systems. Some of the advantage attributed to them in terms of modeling expressivity and analysis techniques apply, at least in principle, to other similar formalisms such as Petri nets for which an approach similar to ours has been developed in \cite{vicario2}, see Section~\ref{sec:related}.} \comment{ which has then to be computed with the aid of numerical or computer algebra methods.} For timed automata, this iterative computation works on \emph{sets} of clock valuations (zones) \cite{hnsy,kronos,uppaal} that each qualitative sequence of events may lead to, where clock values eliminate qualitative behaviors which are infeasible due to timing. In the probabilistic setting, we decorate zones with additional probability information concerning runs and clock values. Thus \comment{, in addition to impossible behaviors,}  we can eliminate classes of behaviors which are feasible but unlikely. Hopefully, the non-negligible overhead associated with computing probabilities will be compensated by the liberty not to explore paths of low probability.

\begin{figure}
\begin{center}
\scalebox{0.7}{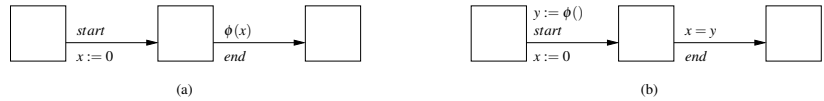 }
\end{center}
\caption{A process that takes time: (a) standard description; (b) decoupling the non-deterministic choice from the \emph{end} transition.  \label{fig:one-step-bis} }
\end{figure}

We consider processes constructed from very simple components such as the automaton of Fig.~\ref{fig:one-step-bis}(a). Such a process is in a waiting state, until it takes a \emph{start} transition and moves to an active state. Clock $x$, which is set to zero upon the transition, measures the time elapsed since the activation. A \emph{start} transition is instantaneous and is initiated by some external scheduler/supervisor. The timing of an \emph{end} transition is based on the clock value and the temporal guard $\phi(x)$, which in the case of timed automata, is simply the condition  $x\in [l,u]$. In duration probabilistic automata we associate a probability density with the duration of each step  which is technically expressed as the distribution over the values of clock $x$ when the \emph{end} transition is taken (note that once started, a process cannot be aborted). We want to analyze the behavior of \emph{multiple} such systems running \emph{concurrently}, each with its own clock.

We use a slightly modified (but equivalent) version of the basic automaton, as shown at Fig.~\ref{fig:one-step-bis}(b). Rather than having the \emph{start} transition deterministic and delegating the non-determinism to the \emph{end} transition, we use an auxiliary variable $y$ which is assigned non-deterministically upon \emph{start} and which should be equal to $x$ upon \emph{end}. In the set-theoretic setting this means an assignment $y\in [l,u]$ while for DPA this means drawing a value for $y$ according to $\phi$, which we denote by $y:=\phi()$.

\comment{In the former case  it does not make a difference but in the latter it helps maintaining probabilistic correctness.}

\begin{figure}
\begin{center}
\scalebox{0.7}{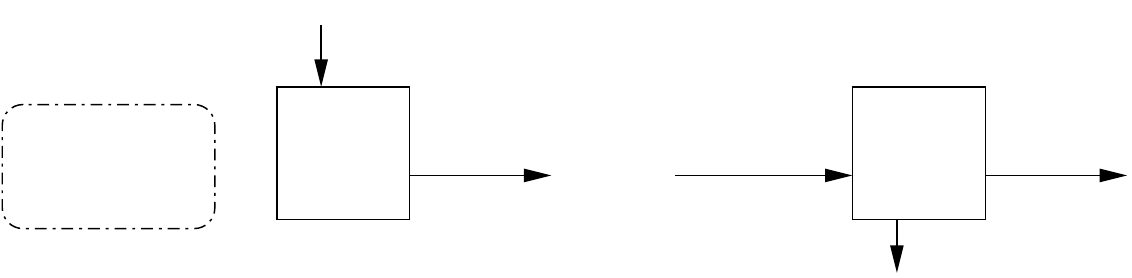 }
\end{center}
\caption{A race. \label{fig:race1-bis} }
\end{figure}

The fundamental phenomenon in the analysis of continuous-time stochastic processes is that of a \emph{race} which occurs in a global state where two or more processes are active. We would like to know which process terminates first, in other words, via which of the pending \emph{end} transitions will the automaton  leave the state. The outcome of a race depends on two factors: the random choices of the respective task durations (the $y$ variables) and  the values of the clocks upon \emph{entering} the global state. Fig.~\ref{fig:race1-bis} shows a fragment of a global automaton representing two parallel processes, both active at state $q$. Clock $x_2$ was reset upon entering $q$, while clock $x_1$, corresponding to a different process that has not yet terminated, was reset in a preceding global state. The gap between the two starting times is maintained by the difference $x_1-x_2$ which remains constant throughout the sojourn in $q$. The larger is this difference, the more likely is clock $x_1$ to satisfy its temporal guard by reaching $y_1$ before $x_2$ reaches $y_2$.\ft{It is interesting to note that in the stochastic processes literature \cite{deds-book}  the role of $x$ and $y$ is taken by a \emph{single} timer $z=y-x$ which is a clock going with derivative $-1$ to zero, after being assigned a random duration. The difference between the two formulations is that ours distinguishes the information that is observable at any time, the value of $x$, from the information that is observed only upon termination, the actual duration $y$. This two-variable representation may provide for more refined \emph{dynamic} schedulers that can base their decisions on the value of $x$, as demonstrated in \cite{schedule-tcs}.}

In the probabilistic setting, this is rephrased as follows. Suppose we  enter a global state with some probability over clock values, and in this state there are several pending \emph{end} transitions guarded by probabilistically-chosen  durations. The probabilities over the clock values upon entrance together with the probabilities over the durations determine the probability that a certain transition wins the race and is taken, as well as the probabilities over the clock values upon taking each of the transitions. We develop a computational scheme for computing for every finite sequence of events the probability that it occurs and the probability over the clock values upon the occurrence of its last event. Technically this is achieved via the concept of a \emph{density transformer} which extends the symbolic  \emph{successor} operator of timed automata (which deserves to be called a subset/zone transformer) to an operator on (partial) densities over clock values. \comment{The main difference between the two is that the probabilistic transformer is measure preserving.}

\section{Definitions}

Throughout this paper we use $\T=[0,\infty)$ as a time domain on which we define probabilities.
We use a fixed set of clock variables $X=\{x_1,\ldots,x_n\}$ all ranging over $\T$ or a bounded subset of it.

\begin{definition}[Clock Constraints and Zones]
The set of clock constraints over $X$,  is defined by the following grammar:
$
\f::=true\,|\, x_i\prec k\,|\, x_i-x_j\prec k\,|\,\f\wedge\f'
$,
where $x_i,x_j\in X$, $k\in\N$ and $\prec\in\{<,\leq,=,\geq,>\}$.
The set of points satisfying a clock constraint is called a zone
\end{definition}
Each zone is a convex polytope in some dimension $m\leq n$ defined
as the intersection of half-spaces which are either orthogonal ($x_i\prec k$) or diagonal ($x_i-x_j\prec k$) with integer $k$. There are finitely many zones in any bounded subset of $\T^n$ or any of its subspaces. We use $\bot$ to denote the zone associated with dimension zero (where no clock is active).

\begin{definition}[Time Densities]
A  piecewise-continuous function $\phi:\T\to \T$ is a time density if it satisfies
$$\int_0^\infty \phi(\tau)d\tau=1.$$
A density has a bounded support $[a,b] \subset \T$ if $\phi(\tau)\not =0 \leftrightarrow \tau\in [a,b]$.
A bounded support  density is uniform if $\phi(\tau)=1/(b-a)$ when $\tau\in [a,b]$.
\end{definition}
The generalization to higher dimension is:
\begin{definition}[Clock Densities]
A function $\psi:\T^m \to \T$ is a clock density if it satisfies
$$\int_0^\infty \ldots \int_0^\infty \psi(\tau_1,\ldots,\tau_m)d\tau_1\ldots d\tau_m=1.$$
We will consider clock densities whose supports are zones.\ft{More precisely, due to resets that put all the probabilistic mass of some clocks at zero, we have to deal with hybrid objects that combine discrete and continuous probabilities and can be framed in terms of densities using impulse functions. \comment{We will not elaborate on this technicality.} }
\end{definition}
Abusing  terminology we call $\psi$ a \emph{partial density} if the above integral is smaller than $1$.

To define the behaviors of our automata we will use timed words (the \emph{time-event sequences} of \cite{acm}) over an alphabet $\S$ of events which will correspond to the various \emph{start} and \emph{end} actions.
\begin{definition}[Timed Words and Languages]
A timed word over a finite alphabet $\S$ is a concatenation of the form $\xi=t_1\cdot w_1 \cdot t_2 \cdot w_2 \cdots$
where $t_i\in \T$ and $w_i\in \S^+$. The untiming of $\xi$ is $\mu(\xi)=w_1\cdot w_2\cdots $ and its duration is  $\l(\xi)=\sum_i t_i$. A timed language is a set of timed words.
\end{definition}

Intuitively this object represents an alternation between  passages of time of duration $t_i$, followed by  sequences $w_i$ of one or more instantaneous events. The events will be \emph{start} and \emph{end} transitions and time passages correspond to time elapsing in active states. All events in  $w_i$ occur at the same absolute time instant $\sum_{j=1}^i t_j$ but in order not to extend the alphabet to $2^\S$ we will consider them as occurring sequentially. We use $\eps$ for the empty word. A timed word $\xi'$ such that $\mu(\xi')=\mu(\xi)$ agrees with $\xi$  on the \emph{order} of events. All such behaviors form an equivalence class $[\xi]$ that we sometime refer to as a \emph{qualitative behavior}. \comment{Given a timed language $L$ and a word $w$ such that $w\in \mu(L)$ we let $L_w=\{\xi \in L: \mu(\xi)=w\}$. Clearly $L$ can be written as
$L=\bigcup_{w\in \mu(L)} L_w$. }

Duration probabilistic automata (DPA)  constitute a well-structured class of timed automata obtained as products of simple DPA and a scheduler. They can model most situations encountered in the analysis of scheduling problems such as job-shop or task-graph \cite{schedule-tcs} and are free from notorious anomalies such as Zeno behaviors.
For economy of expression, we use as our building blocks processes that admit several processing steps where the \emph{end} transition of step $j$ leads to the waiting state of step $j+1$. Although practically, the same clock can be reused in subsequent steps, conceptually we prefer sometimes to view each step $j$  as using a distinct clock $x^j$. \comment{ associated with a distinct random variable.} Let $N=\{1,\ldots,n\}$ and $K=\{1,\ldots,k\}$.

\begin{figure}
\begin{center}
\scalebox{0.6}{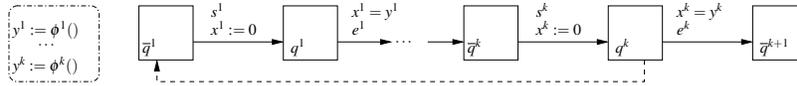 }
\end{center}
\caption{A simple DPA: acyclic and cyclic (dashed transition).\label{fig:simple-new} }
\end{figure}

\begin{definition}[SDPA]
A simple duration probabilistic automaton (SDPA)  of $k$ steps is a tuple $\A=(\S,Q,X,Y,\D,\wq^1)$ where $\S=\S_\st\uplus \S_e$ is the alphabet of \emph{start} and \emph{end} actions with  $\S_\st=\{s^1,\ldots, s^k\}$ and
$\S_e=\{e^1,\ldots, e^k\}$.  The state space is an ordered set $Q=\{\wq^1,q^1,\wq^2,\ldots, q^{k},\wq^{k+1}\}$ with $\wq^j$ states considered \emph{idle} and $q^j$ states are \emph{active},  $X=\{x^1,\ldots,x^k\}$ is a set of clock variables and $Y=\{y^1,\ldots,y^k\}$ is a set of auxiliary random variables, each distributed according to a bounded and uniform time density $\phi^j$. The transition relation $\D$ consists of two types of transitions:
 \begin{enumerate}
 \item \emph{Start} transitions:  for every idle state $\wq^{j}$, $j\in K$, there is one transition of the form $(\wq^{j},\st^j,\{x^j\},q^{j})$. When the transition is taken,  clock $x^j$  is reset to zero and becomes active. Such  transitions take no time;
 \item \emph{End} transitions: for every active state $q^{j}$, $j\in K$,  there is a transition of the form $(q^{j},x^j=y^j,e^j,\wq^{j+1})$. This transition renders clock $x$ inactive.
\end{enumerate}
State $\wq^1$ is the  initial state of $\A$.
\end{definition}
The SDPA just defined is acyclic. A cyclic version of this definition, employs addition  modulo $k$ with the last transition going back to $\wq^1$, see Fig.~\ref{fig:simple-new}. \comment{Computing the long-term behavior of cyclic automata requires a fixed-point theory for the operators defined} In this paper we restrict ourselves to acyclic automata.
%

The operational interpretation is the following: for each step $j$ we draw a duration $y^j$ according to $\phi^j$. \comment{ and this value determines the duration of the step.} Inside an active state $q^{j}$, clock $x^j$ advances with derivative $1$ and the \emph{end} transition is taken when $x^j=y^j$, that is, $y^j$ time after the corresponding \emph{start} transition. A generalized state (configuration) of the automaton in an active state is a pair $(q,v)$ consisting of a discrete state and a clock value $v$ which represents the time elapsed since the last \emph{start} transition. Note the difference between transition labels $\st^j$ and $e^j$: the former is an external command coming from a \emph{scheduler} outside the SDPA, while the latter is emitted by the SDPA itself when it terminates a step within a randomly chosen duration. When such a scheduler is not specified, the automaton can be viewed as non-deterministic, generating behaviors of the form
$$r^1\cdot s^1\cdot t^1 \cdot e^1\cdot r^2 \cdot s^2 \cdot t^2 \cdot  e^2 \cdots \cdot s^k\cdot t^k \cdot e^k \cdot \infty$$
with each $r^j\in \T$ being an arbitrary waiting period and each $t^j$ is in the support  of $\phi^j$.

Duration probabilistic automata (DPA) are obtained by composing a set  $\{\A_i=(\S_i,Q_i,X_i,\D_i,\wq^1_i)\}_{i\in N}$ of SDPA  with a \emph{scheduler}. To simplify notations we assume all $\A_i$ to admit the same number $k$ of steps. The event alphabet is the union of the event alphabets $\S_i$,
that we write as  $\S=\S_\st\uplus \S_e$ with $\S_\st=\{\st_i^j:i\in N,j\in K\}$ and $\S_e=\{e_i^j:i\in N,j\in K\}$. The
state space of the product automaton is  $Q=Q_1\times\cdots\times Q_n$. The composition of automata, which is  fairly standard in the non-deterministic setting, often employs an  \emph{interleaving semantics} where independent transitions can occur in any order. Applying this approach to several \emph{start} transitions that take place \emph{simultaneously}, introduces an annoying artificial non-determinism that we avoid by combining all transitions that occur simultaneously into a single transition. However in order to maintain the semantics of the automaton as a set of timed words over $\S$ we will associate with such a transition a unique \emph{sequence} of labels. This is done via a \emph{sequentialization function} which maps every $E\se \S$ into a sequence $\a(E)\in \S^+$ consisting of the elements of $E$ concatenated according to some fixed order relation over the alphabet. We say that transition $s_i^j$ is \emph{enabled} in global state $q$ if the $i^{th}$ component of  $q$ is $\wq_i^j$.

\begin{definition}[Scheduler]
A scheduler for a set $\{\A_i\}_{i\in N}$ of SDPA  is a function $S:Q\to 2^{\S_\st}$, satisfying:
 \begin{itemize}
 \item $s_i^j\in S(q)$ only if  $s_i^j$ is enabled in $q$;
  \item $S(q)=\emptyset$ only if $q$ is the global final state or admits at least one active component.\comment{\ft{ We assume that schedulers are monotone non-increasing with respect to the number of active components, that is, if the set of active processes in $q'$ is a superset of  those active in $q$ then $S(q')\se S(q)$. This prevents several sequences of \emph{start} transition to be taken without an \emph{end} transition between them.}}
  \end{itemize}
\end{definition}

The scheduler plays two roles in our model. First, it guarantees mathematical sanity with a single run for every value of the random variables and a non-blocking behavior where all prefixes of runs have continuations that reach the final state in a bounded amount of time. In a world of unlimited resources where each SDPA may progress independently, $S(q)$ is the set of all transitions enabled in $q$ and the scheduler is restricted to this mathematical role.  The more interesting case is when the scheduler has to resolve resource conflicts and keep some processes waiting while giving priority to others. Abusing notation we  say that $i\in S(q)$ if $s_i^j\in S(q)$ for some $j$.

\comment{A \emph{scheduler} is a device which for every global state $q=(q_1,\ldots,q_n)$ decides which subset of the enabled \emph{start} transitions is taken. Formally it is defined as .}

\begin{definition}[Duration Probabilistic Automata]
A duration probabilistic automaton (DPA) is a composition $\A=\A_1\circ \cdots \circ \A_n\circ S=(Q,X,Y\D,q^0)$ of $n$ SDPA and a scheduler. The state space is $Q\se Q_1\times\cdots Q_n$  with initial state $q^1=(\wq_1^1,\ldots,\wq_n^1)$  the set of clocks\ft{Since at any time there is at most one clock active for each $\A_i$, we will sometimes refer to the set of clocks as $\{x_1,\ldots,x_n\}$ where $x_i$ refers to some $x_i^j$ depending on the state of $\A_i$. Likewise we will compare it with $y_i$ denoting the appropriate $y_i^j$.} is  $X=\bigcup_i X_i$ and the auxiliary variables $Y=\bigcup_i Y_i$.   The transition relation $\D$ consists of two types: multiple \emph{start} transitions of the form $(q,w,R,q')$ where $w\in \S_{\st}^+$ is a sequence of labels and $R$ is a set of initialized clocks, as well as \emph{end} transitions of the form $(q,x_i=y_i,e_i,q')$, one for each $\A_i$  active in $q$.
 \begin{itemize}
  \item For every state $q=(q_1,\ldots,q_n)$ such that $S(q)=E\not=\emptyset$ we define a transition
$$((q_1,\ldots,q_n),w,R,(p_1,\ldots,p_n))$$
where $w=\a(E)$ is the sequentialization of $E$ and $R=\{x_i^j:s_i^j\in E\}$. When $i\not\in E$  $p_i=q_i$ otherwise $p_i=q'_i$  where $(q_i,s_i,\{x_i\},q'_i)\in \D_i$ is the corresponding \emph{start} transition;
\item For every $q$ such that $S(q)=\emptyset$ and for every $i$ such that $q_i$ is active and $(q_i,x_i=y_i,e_i,q'_i)\in\D_i$ is its corresponding \emph{end} transition, we define a transition
$$((q_1,\ldots,q_i,\ldots,q_n),x_i=y_i,e_i,(q_1,\ldots,q'_i,\ldots,q_n))$$
\end{itemize}
\end{definition}
This definition gives priority to the immediate \emph{start} transitions while the pending \emph{end} transitions are allowed only in a state where no immediate transitions are admitted by the scheduler.



\section{Behaviors and their Computation}

 The set of all complete behaviors that a DPA $\A$ may generate constitutes a timed language $L=L(\A)$. The  probabilistic semantics of $\A$ is a probability distribution over subsets of $L$. We will not give at this point a detailed formal definition of this semantics but rather convey sufficient intuition to relate it to the zone-based computation that we develop in the sequel. For the sake of simplicity, we temporarily assume a most liberal scheduler which executes every  $s^j_i$ immediately after $e_i^{j-1}$. The untiming $\mu(L)$ of the language consists of words satisfying some well-formedness condition, that is, $\mu(L)\se M$ where  $M=M_1||\cdots||M_n$ is the shuffle of the SPDA local languages, each of the from $M_i= \{s_i ^1\cdot e_i ^1 \cdots s_i^k\cdot e_i^k\}$.
By construction, there is a one-to-one correspondence between sequences of events in $M$ and complete paths in $\A$. Hence $L$ can be written as a union $\bigcup_{w\in M} L_w$ of languages, each corresponding to a subset of $L$ corresponding to a particular \emph{order} of events. Elements of $L_w$ are obtained from $w$ by inserting time durations between the events.

Each choice $y$ of values for the duration random variables determines a \emph{unique} behavior of the system
that we denote $\xi(y)$ and the probability of a set of behaviors is the probability of the $y$ values that induce them. The density of this distribution at a complete timed word $\xi=t_1 \s_1,\ldots,t_{nk} \s_{nk}$ under a liberal scheduler is defined as follows.
For every step $(i,j)\in N\times K$, let $r_i^j$ be the sum of all duration occurring between $s_i^j$ and $e_i^j$. Then the density at  $\xi$ is:
\begin{equation}
\label{eq:complete}
\prod_{i\in N,j\in K} \phi_i^j(r_i^j).
\end{equation}
Unfortunately (\ref{eq:complete}) cannot be exported as is to the case of non-trivial schedulers where we have to resort to \emph{incremental} computations that derive the probability of $\xi\cdot t \cdot \s$  from the probability of
its \emph{prefix} $\xi$. To this end we need to consider \emph{incomplete} behaviors that correspond to a word $w$ in which \emph{not every} $s$ has been followed by a matching $e$.
The probability of $\xi\cdot t\cdot e$ for each of the pending \emph{end} events depends on the probability of  the corresponding step to terminate within a duration equal to the sum of $t$ and the duration in $\xi$ occurring after $s$ \emph{and} the probability of the \emph{other} steps already started in $w$ to terminate \emph{after} that.

\comment{All these depend both on the duration distributions of the corresponding tasks, and the distribution of the time elapsed since their corresponding \emph{start} transitions, information that timed automata keep as clocks.}

An incomplete behavior $\xi$ can be associated with two other objects, the first being the subset of $L$ consisting of complete words having $\xi$ as a \emph{prefix} and the second is a global configuration of the automaton reached while generating $\xi$. A global state in a timed automaton is a mixture of active and idle local states with active clocks defined naturally according to the state, and this determines the dimensionality of clock space in that state. Thus a configuration is  a pair $(q,v)$ with $v\in\T^m$ for some $m\leq n$, and the time evolution inside the state consists of all active clocks advancing in the same pace, keeping the \emph{difference} between any pair of active clocks  constant throughout the sojourn in a state. The set of \emph{time predecessors} of a clock valuation  is
$\tpr(v)=\{v-\t\vi: \t\geq 0\}\cap \T^m$,
where $\vi$ is a vector $(1,\ldots,1)$ of dimension $m$. A configuration $(q,v)$ can be reached via time passage \emph{only} from configurations of the form $(q,v')$ with $v'\in \tpr(v)$.

\begin{figure}
\begin{center}
\scalebox{0.65}{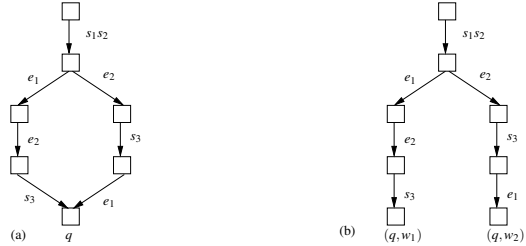 }
\end{center}
\caption{(a) Two commuting paths; (b) splitting a state into two copies according to the history.\label{fig:commute1}  }
\end{figure}

Let us just comment on the issue of commuting paths in the automaton. Why can we  merge two such paths into a single state despite their differing past histories? The reason is that the past events that occurred in different orders along the two paths are of two types: 1) events related to \emph{completed} steps that do not affect the future beyond what is already encoded in the state; 2) \emph{start} transitions of steps which are still active in $q$. These events do affect the future but the order of their occurrence is captured already, at a finer level of detail, by the values of the active clocks and their pairwise \emph{differences}. This is illustrated in the two commuting paths depicted in Fig.~\ref{fig:commute1}(a), assuming step $3$ to follow step $2$ in the same SDPA. The qualitative languages associated with the paths are the singletons $w_1=s_1 s_2 e_1 e_2 s_3$ and $w_2=s_1 s_2 e_2 s_3 e_1$, respectively, while the qualitative language of the whole state $q$ is $s_1 s_2 (e_1|| e_2 s_3)$ and the only information that still affects the future is the time elapsed since $s_3$, captured by a clock (see also \cite{interleave}). Despite this fact, for convenience reasons, we split states according to their respective histories, that is, work with \emph{extended discrete states} of the form $(q,h)$ where $h\in \S^*$. A transition from $q$ to $q'$ labeled by some $w\in \S^+$ thus extends into a transition from $(q,h)$ to $(q',h\cdot w)$, and the transition graph of the automaton becomes a tree, see Fig.~\ref{fig:commute1}(b).

We will present our method to compute the probabilistic semantics gradually starting  with its \emph{support}, which is the set of all  timed words  which are possible if we interpret each $\phi$ as an interval, as in timed automata. Although what is described in the sequel is standard material underlying the \emph{practice} of TA verification tools \cite{kronos,uppaal}, it is our perception that it is not sufficiently known to the more general public. \comment{, and certainly not among those working on stochastic processes.}  We assume that for every component $i$ active in state $q$, the duration of its corresponding step is distributed with a uniform density $\phi_i$ of support $[a_i,b_i]$. We use $R(v)$ to denote the setting to zero of clocks in $R$ and the continuation of the clocks in $v$ that are not in $R$. Note that by the definition of SPDA all clocks in $R$ are inactive in $q$ before the transition.


\begin{definition}[Steps and Runs]
 A step of a DPA $\A$ is one of the following:
\begin{itemize}
\item A \emph{start step}: $(q,h,v)\stackrel{{w}}{\longrightarrow}(q',h\cdot w,v')$,
for some  $(q,w,R,q')\in\Delta$ such that $v'=R(v)$;
\item A \emph{time step}: $(q,h,v)\stackrel{{\t}}{\longrightarrow}(q,h,v+\t\vi)$;
for some $\t>0$ such that for every $i$ active in $q$, $v_i+\t\leq b_i$; \comment{Before I had the extra condition that  for some $i$, $v_i+\t\leq a_i$ in order to guarantee that after the time step you can make a transition and do not make redundant time steps, but I don't think it is needed (and it cannot prevent stuttering in time).}
\item An \emph{end step}: $(q,h,v)\trans{e_i} (q,h\cdot e_i,v')$ where $v_i\in [a_i,b_i]$ and $v'$ is obtained from $v$  by deactivating $x_i$.
\end{itemize}
A run of the automaton is a sequence of steps which starts at $(q^1,\eps,\bot)$  and alternates between single time steps and one or more transition steps.
\end{definition}
The behavior associated with a run is the timed word obtained by concatenating the labels (transitions and durations) of its steps. We use the notation $(q,h,v)\trans{\xi} (q',h',v')$ to denote a run from $(q,h,v)$ to $(q',h',v')$ generating the timed word $\xi$ (note that $h'=h\cdot \l(\xi)$). We also use the notation $(q,h,v)\trans{\xi} (\cdots) $ to denote an infinite run starting from $(q,h,v)$. For acyclic DPA, all such runs terminate with an infinite time step inside the final state.
\begin{definition}[State Languages]
With every extended configuration $(q,h,v)$ we associate the following timed languages:
\begin{itemize}
\item The set of  behaviors associated with runs whose last event is a $\s$-labeled transition to $(q,h\cdot \s)$:
$$\ltq(q,h,v)=\{\xi\cdot \s:(q^1,\eps,\bot)\trans{ \xi \cdot \s }  (q,h\cdot \s,v)\}$$
\item The infinite behaviors generated by runs that start from $(q,h,v)$:
$$\lfq(q,h,v)=\{\xi:(q,h,v)\trans{\xi} (\cdots)\}$$
\item The set of all infinite behaviors of $\A$ with prefixes in $\ltq(q,h,v)$:
$$L(q,h,v)=\ltq(q,h,v)\cdot \lfq(q,h,v).$$
\end{itemize}
\end{definition}
\begin{observation}
\label{obs}
If $\xi\in \ltq(q,h,v)$ then $v_i$ is equal, for every $i$ active in $q$, to the time elapsed since the last $s_i$ event in $\xi$.
\end{observation}

\begin{definition}[Symbolic States]
An (extended) symbolic state is a triple $(q,h,Z)$ with $q\in Q$, $h\in \S^*$ and $Z$ is a zone of dimensionality compatible with $q$.
\end{definition}
Intuitively, $Z$ will be the set of all possible clock values  that runs along the path to $(q,h)$  may have. We will lift the definition of state languages to symbolic states by letting  $L(q,h,Z)=\bigcup_{v\in Z} L(q,h,v)$.
We  associate with time passage and with every transition a successor operator over symbolic states.
\begin{definition}[Successor Operator]
Successor operators admit three types:
\begin{itemize}
\item Time successors: $post^t(q,h,Z)=(q,h,Z')$ where $$Z'=\{v':\exists v\in Z\,\exists \t\in\T ~(q,h,v)\trans{\t} (q,h,v')\}$$
\item Start successors: $post^s(q,h,Z)=(q',h\cdot w,R(Z))$ for every \emph{start} transition \\ $(q,w,R,q')\in \D$;
\item End successors: $post^e(q,h,Z)=(q',h\cdot e,Z')$ for every transition $(q,x=y,e,q')\in \D$ where $Z'$ is obtained from $Z$ by eliminating the appropriate de-activated clock.
\end{itemize}
\end{definition}
The reachability graph, also known as the simulation graph, is what timed automata verification tools \cite{kronos,uppaal} compute as a symbolic representation of the semantics of the automaton.
\begin{definition}[Reachability Graph]
The reachability graph associated with a DPA $\A$ is a graph of symbolic states obtained by successive application of successor operators to $(q^1,\eps,\bot)$.
\end{definition}
The fundamental property of the reachability graph is the following.
\begin{theorem}
A symbolic state $(q,h,Z)$ is part of the reachability graph iff for every $v\in Z$, the language $\ltq(q,h,v)$ is not empty.
\end{theorem}
In other words there is a timed word $\xi$ generated by the automaton with $\mu(\xi)=h$ such that for every active component $\A_i$, the duration of the suffix of $\xi$ starting with the last $s_i$ event  is $v_i$. Note that since all runs of a DPA have a continuation, $\ltq(q,h,v)\not=\emptyset$ implies that $\lfq(q,h,v)\not=\emptyset$ and $L(q,h,v)\not=\emptyset$. Moreover, $L(q,h,Z)$ is exactly $L_h$, the set of timed words in $L$  whose untiming is $h$.  \comment{$L_h$ of Definition~\ref{def:singular}}

In the sequel we will extend the reachability graph with probabilities and work with symbolic states of the form $(q,h,Z,\psi)$ where $\psi$ is a partial density function over the clock values in $Z$, which can be used to compute  the probability of $L(q,h,Z)$ or its subsets. To this end we need to extend the successor operators to become density transformers.

\comment{
Note that the concatenation of two time steps of durations $d_1$ and $d_2$ is a time step of duration $d_1+d_2$. \comment{and, conversely, \comment{due to the dense nature of the real numbers,} a time step can split into any number of smaller steps.}
A \emph{compound step} is a discrete step  followed by a time step (possibly
of zero duration):
$$(q,v)\stackrel{{\d,d}}{\longrightarrow}(q',v'+d)\equiv(q,v)\stackrel{{\d}}{\longrightarrow}(q',v')\stackrel{{d}}{\longrightarrow}(q',v'+d)$$
A \emph{run} of the automaton $\mathcal{A}$ starting from a configuration $(q,v)$ is a sequence of compound
steps.\comment{The first step can be a pure time step.} We use the notation $(q,v)\stackrel{{\xi}}{\longrightarrow}(q',v')$ for runs. The untiming of $\xi$, denoted by $\mu(\xi)$, is
the sequence of transitions taken, regardless of the time durations. The semantics of $\A$, denoted by $\sem{\A}$, is the set of runs  it may generate from the initial state and its qualitative semantics is $\mu(\sem{\A})=\{\mu(\xi)|\xi\in\sem{\A}\}$.
}

\section{Density Transformers}

\begin{figure}
\begin{center}
\scalebox{0.7}{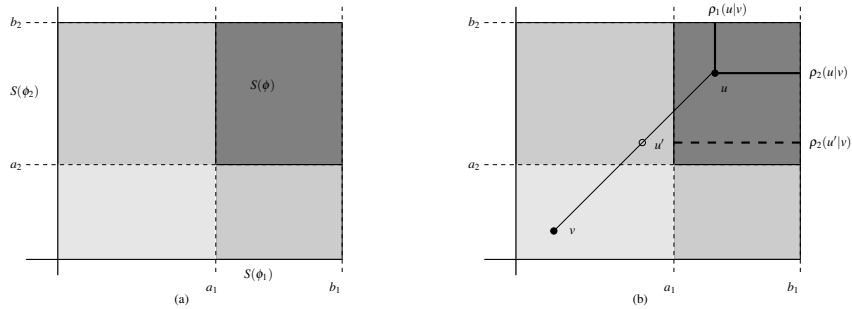 }
\end{center}
\caption{A race: (a) a state can be entered at any point in the shaded area; transition can be taken only in the darker area; (b) the probabilities $\rho_1$ and $\rho_2$. \label{fig:dense}  }
\end{figure}

The major issue in our computational approach is to determine, in a state where several processes are active, the probability of each of  the pending \emph{end} transitions to be taken and how the clock values are distributed when the transition is taken. As an informal illustration consider state $q$ in the automaton of Fig.~\ref{fig:race1-bis}  admitting two competing active processes whose durations are distributed with densities $\phi_1$ and $\phi_2$, respectively. Assuming both $\phi_1$ and $\phi_2$ are uniform with a bounded interval support, their joint density $\phi(y_1,y_2)=\phi_1(y_1)\phi_2(y_2)$ is supported by a rectangle of the form $[a_1,b_1]\times [a_2,b_2]$. The clock values with which the state can be entered are restricted to the rectangle $[0,b_1]\times [0,b_2]$ and the two transitions can be taken in the rectangles $[a_1,b_1]\times [0,b_2]$ and
$[0,b_1]\times [a_2,b_2]$, respectively, see Fig.~\ref{fig:dense}(a). Note that the points of exit need not be inside the (joint) support of $\phi$.

What is the probability $\rho_i(u|v)$ that  transition $i$ is taken at some point $u=(u_1,u_2)$, i.e., $u_i=y_i$, \emph{given} that the state has been entered at some $v$? First of all, this probability is non-zero only if $v\in\tpr(u)$, that is, $v$ is a time-predecessor of $u$. Secondly,
for transition $1$ to be taken, it should be the case that process $1$  chooses duration $u_1$ while process $2$  chooses some $y_2> u_2$ (the vertical thick line in Fig.~\ref{fig:dense}(b)). Transition $2$ will be taken at $u$ when process $2$ chooses a duration $u_2$ and process $1$ some  $y_1> u_1$ (the horizontal thick line in the figure). Thus $\rho_1(u|v)$ is obtained by summing up the duration probabilities \emph{above} $u$ and $\rho_2(u|v)$ by summing up the probabilities \emph{to the right} of $u$. Note that $\rho_i(u|v)=\rho(u|v')$  for any other $v'\in\tpr(u)$ and that for points like $u'$ outside the support of $\phi_1$ we will have $\rho_1(u'|v)=0$ and $\rho_2(u'|v)=1$. Assuming that the state has been entered with some density $\psi$ over clock values, we can sum up $\rho_i(u|v)$ over $v\in \tpr(u)$ according to $\psi$ and obtain the expected $\rho_i(u)$ as well as new densities $\psi_i$ reflecting the distribution of the clock values upon taking each of the transitions.

With every extended state $(q,h)$ in which $m$ processes are active we associate a partial density function of the form $\psi(x_1,\ldots,x_m,y_1,\ldots,y_m)$ whose intended meaning is to capture the probability over clock values upon \emph{entering} the state. Although the $y$ variables are static and do not vary during execution, we need to keep them in the picture because they do not distribute evenly as time goes by. In other words, certain combinations of choices of durations will make some transitions impossible. We associate density transformers with every \emph{start} and \emph{end} transition as follows.

\ni {\bf Start}: Let $q$ be a state with $l$ active components and let $s$ be a \emph{start} transition which activates processes $\{l+1,\ldots,m\}$.\ft{The restriction to these indices is just to simplify notation. Recall also our previous remark that our probabilities are in reality hybrid, mixing discrete probabilities and distributions.} We associate with $s$ the density transformer $\dt_s$ such that $\psi'=\dt_s(\psi)$ if
$$\begin{array}{lcr}\psi'(x_1,\ldots,x_l,0,\ldots,0,y_1,\ldots,y_l,y_{l+1},\ldots,y_m)  & = &
\psi(x_1,\ldots,x_l,y_1,\ldots,y_l)\\ & & \cdot \phi(y_{l+1},\ldots,y_m)
\end{array}
$$
with $\phi(y_{l+1},\ldots,y_m)=\phi_{l+1}(y_{l+1})\cdots \phi_m(y_m)$.
When one of $\{x_{l+1},\ldots,x_m\}$ is non-zero, $\psi'=0$. This operation just reflects the setting  of the new clocks to zero and the introduction of their respective durations.

\ni{\bf End:} For every \emph{end}  transition $e_i$ outgoing from a state $q$ with $m$ active processes  we define two density transformers $\dt_{r_i}$ and $\dt_{\bot_i}$.
As explained previously, the transformer $\dt_{r_i}$  computes the clock density at the time when process $i$ wins the race, given the density was $\psi$ upon entering the state. It is defined as $\psi_i=\dt_{r_i}(\psi)$ if
$$
\begin{array}{l}
\psi_i(x_1,\ldots,x_m,y_1,\ldots,y_m)= \\ \\
~~~~~~~~~~~~\left\{\begin{array}{rl} { \ds \int_{\t>0}\psi(x_1-\t,\ldots,x_m-\t,y_1,\ldots,y_m)d\t } &
\mbox{~~~if~}x_i=y_i \wedge \forall i'\not=i~ x_{i'}<y_{i'} \\ \\
0 & \mbox{~~~otherwise}
\end{array}\right.
\end{array}
$$
The transformer  $\dt_{\bot_i}$, which  just deactivates clock $x_i$ and projects it away from the clock space is defined as $\psi'=\dt_{\bot_i}(\psi)$ if
$$
\begin{array}{l}
\psi'(x_1,\ldots,x_{i-1},x_{i+1},\ldots, x_{m},y_1,\ldots,y_{i-1},y_{i+1},\ldots,y_{m})= \\
~\\
~~~~~~~~~~~~~~~~~~~~~~~~~~~~~~~~~~~~~~~~~~~~~~~~\ds{\int_{u_i}} \psi(x_1,\ldots,u_i,\ldots,x_m,y_1,\ldots,u_i,\ldots,y_m)d u_i.
\end{array}
$$

We can now define a probabilistic version of the successor operators. Note that for timed automata we had a unique time successor operator for each state, while for DPA time successor operators are specific for each of the transitions that participates in the race. \comment{\ft{Explain why it is required by probabilistic correctness?}} A probabilistic symbolic state is a tuple $(q,h,Z,\psi)$.

\begin{definition}[Probabilistic Successor Operator]
Probabilistic successor operators admit two types:
\begin{itemize}
\item Start successors: $post^s(q,h,Z,\psi)=(q',h\cdot w,R(Z),\psi')$ for every \emph{start} transition  $(q,w,R,q')\in \D$ where $\psi'=\dt_s(\psi)$;
\item End successors: $post^{e_i}(q,h,\psi,Z)=(q',h\cdot e_i,Z',\psi')$ for every transition $(q,x=y,e_i,q')\in \D$ where $\psi' =\dt_{\bot_i}(\dt_{r_i}(\psi))$ and $Z'$ is the support of $\psi'$.
\end{itemize}
\end{definition}
The \emph{probabilistic  reachability graph} is computed by starting with the initial probabilistic symbolic state $(q^1,\eps,\bot,\psi_\bot)$  and then applying the appropriate successor operators. Computing this graph, as in the case of timed automata, allows us to compute everything of interest for DPA as we show below.

Recall that every $y$ valuation induces a complete run $\xi(y)$ with an untiming $h$. Grouping all the $y$ values resulting in the same $h$ we have a mapping from the duration space $\R^{nk}$ to the finite set $\S^{nk}$ which defines the probability of each path. To extend this notion to incomplete behaviors one could define a sequence of functions $\{f_\a:\a=0,\ldots,nk\}$ over the duration space, each mapping $y$ into a prefix  $\xi(y)_\a$  of $\xi(y)$  admitting \emph{exactly} $\a$ discrete transitions. As mentioned earlier, to compute $f_{\a+1}$ from $f_\a$ it is sufficient to know the qualitative prefix $h_\a$ and the time elapsed since the non-terminated \emph{start} events. For each $\a$ we have then a \emph{hybrid} (discrete-continuous)  probability distribution on $\S^\a\times\R^m$ which can be expressed as a finite set of densities $\{\eta_h:h\in \S^\a\}$. Our main claim is that if $(q,h,Z,\psi)$ is part of the probabilistic reachability graph then $$\eta_h=\int_y \psi(x,y)d y.$$
This holds trivially for the root  $(q^1,\ee,\bot,\psi_\bot)$ which corresponds to $\eta_\ee$ where all the probability is concentrated in the empty sequence. The inductive step, showing that if a node $(q,h,Z,\psi)$ satisfies
$\eta_h(x)=\int\psi(x,y)dy$ than any successor $(q',h',Z',\psi')$ satisfies
$\eta_{h'}(x)=\int\psi'(x,y)dy$, is immediate for a \emph{start} successor because it just concatenates some $s$-labels without changing probabilities. For an \emph{end} successor $e_i$, observe that for every $v\in Z$ and $v'\in Z'$, the corresponding run leads from $(q,h,v)$ to $(q',h\cdot  e_i,v')$, concatenating to the language a timed word $\t\cdot e_i$ with $\t=v'_i-v_i$ and the probability of $v'$, the time elapsed since the remaining uncompleted \emph{start} events, is captured by $\psi'$.

\comment{
\begin{theorem}
If $(q,h,Z,\psi)$ is part of the probabilistic reachability graph then for every set $V$ of clock valuations  the probability of $L(q,h,V)$ among the behaviors of $\A$ is $$\int_{x\in V} \int_y \psi(x,y)dx dy.$$
\end{theorem}
\ni{\bf Sketch of Proof}:
For the base case $L(q^1,\ee,\bot)$ has probability $1$ since this is the set of all behaviors of $\A$. Now suppose this holds for a node $(q,h,Z,\psi)$ in the graph and show it holds for any successor $(q',h',Z',\psi')$ . For a \emph{start} successor this is immediate because it just concatenates some $s$-labels without changing probabilities. For an \emph{end} successor $e_i$, observe that for every $v\in Z$ and $v'\in Z'$, the corresponding run leads from $(q,h,v)$ to $(q',h\cdot  e_i,v')$, concatenating to the language a timed word $\t\cdot e_i$ with $\t=v'_i-v_i$ and the probability of $v'$ is captured by $\psi'$. \qed
}
\comment{A symbolic state $(q,h,Z,\psi)$  gives us all the information about the probability of being in the untimed extended state $(q,h)$ (and its untimed language $\{h\}$) by letting
 $$P((q,h))=\int_x \int_y \psi(x,y)dx dy.$$}

Thus we can compute the probability for each interesting set of paths, for example those in which some event precedes another. Moreover, by adding an auxiliary clock which is never reset and measures absolute time, we can retrieve the evolution of these probabilities over time and compute the distribution and expected value of the total termination times. This provides for an effective comparison between the performance of different scheduling policies.

 \comment{Note also that, like in the case of timed automata where the reachability graph can be used to construct an untimed automaton (time-abstract simulation quotient) admitting the same untimed language, we can build a discrete-time Markov chain whose state are extended states of the form $(q,h)$ with the transition probability between $(q,h)$ and its successor $(q',h')$ is
 $T((q,h),(q',h'))=P((q',h'))/P((q,h))$.
}

 \comment{Note also that it is straightforward to implement an algorithm for verification of DPA up to probability $1-\varepsilon$: just explore the reachability graph and discard nodes of low probability as long as the sum of their probabilities is less than $\varepsilon$.}

\section{Past and Future Work}
\label{sec:related}

We have shown how timed automata verification techniques can be extended to handle durations which are distributed probabilistically. We conclude by mentioning some related work as well as some of the many open issues that remain.

 \comment{Hopefully, framing the problem in terms of automata and timed languages will facilitate the development of practical methods for verification and performance analysis of real-time systems under \emph{realistic}, rather than \emph{pessimistic} assumptions.}

The works closest to ours are those of Alur and Bernadsky \cite{bernadsky1,bernadsky2} and Vicario et al.\ \cite{vicario0,vicario1,vicario2}, each using a different models. The work of \cite{bernadsky1,bernadsky2} is concerned with verifying temporal properties for some classes of GSMPs, where the hard part is the treatment of the unbounded \emph{until} operator which is achieved by putting restrictions on the number of concurrently active clocks. They also deal with computational issues related to symbolic computation of integrals over exponential-polynomial distributions. The work of \cite{vicario0,vicario1,vicario2} is concerned with  certain classes of stochastic Petri nets for which they develop a computational framework similar to ours which includes both exact and approximate computation of the distributions. The major difference is that our formulation that separates the $x$ and $y$ variables, provides for more sophisticated scheduling policies, such as those described in \cite{schedule-tcs}, that take clock values into consideration.

The most urgent topics in our agenda are the implementation and the extension to cyclic DPA. From a computational standpoint, since we start with uniform distribution, all our density transformers result in piecewise-polynomial functions that can be computed analytically using a mixture of zone-based algorithms and computer algebra tools. Of course, the obtained  expressions will become increasingly complex due to case splitting and may require approximation. An alternative (but not scalable) way would be to work using  discrete-time approximations of the duration distributions. The present results allow us to compute reachable symbolic states forward to any desired horizon, but since densities are much richer than zones, there is no immediate proof of convergence to a fixed point. Given that the density transformer can be phrased as a linear operator over state-related densities, we intend to investigate functional analysis techniques like those used in \cite{aldric} to establish convergence and approximate termination.

\ni {\bf Acknowledgment}: This work benefitted from discussions with E.~Asarin and from numerous anonymous referees.

\bibliographystyle{plain}
\bibliography{bib-dpa}

\end{document}